# Structural attributes of nucleotide sequences in promoter regions of supercoiling-sensitive genes: how to relate microarray expression data with genomic sequences

Galina I. Kravatskaya\*, Vladimir R. Chechetkin, Yury V. Kravatsky and Vladimir G. Tumanyan

*Engelhardt Institute of Molecular Biology of Russian Academy of Sciences, Vavilov str., 32, Moscow, Russia 119991*

\*Correspondence: Galina I. Kravatskaya. E-mail: gk@eimb.ru

# Abstract

The level of supercoiling in the chromosome can affect gene expression. To clarify the basis of supercoiling sensitivity, we analyzed the structural features of nucleotide sequences in the vicinity of promoters for the genes with expression enhanced and decreased in response to loss of chromosomal supercoiling in *E. coli*. Fourier analysis of promoter sequences for supercoiling-sensitive genes reveals the tendency in selection of sequences with helical periodicities close to 10 nt for relaxation-induced genes and to 11 nt for relaxation-repressed genes. The helical periodicities in the subsets of promoters recognized by RNA polymerase with different sigma factors were also studied. A special procedure was developed for study of correlations between the intensities of periodicities in promoter sequences and the expression levels of corresponding genes. Significant correlations of expression with the AT content and with AT periodicities about 10, 11, and 50 nt indicate their role in regulation of supercoiling-sensitive genes.

# Introduction

Dynamic alteration of chromosomal supercoiling induces variations of the torsional tension in the DNA double helix and affects the biological processes that involve unwinding of DNA such as replication initiation or transcription [1–4]. A change in expression of supercoiling-sensitive genes (SSG) provides a flexible response to altered nutritional and environmental conditions. This mechanism was studied in detail in the model organism *Escherichia coli* [5–11]. Despite the progress in this field, the molecular mechanisms of switching SSG and their relationship with corresponding DNA sequences are still poorly understood.

Peter et al. [8] used microarrays representing nearly the entire genome of *E. coli* MG1655 and identified statistically significant changes in expression of 306 genes (about 7% of the genome) under relaxation of negative supercoiling. The expression of 106 genes increased upon chromosome relaxation (relaxation-induced genes or RIG), whereas the expression of 200 genes decreased (relaxation-repressed genes or RRG). Peter et al. [8] found that upstream and coding sequences for RIG were AT-rich, whereas the corresponding sequences for RRG had GC preference. They also suggested that the promoters and their interactions with DNA-binding proteins are responsible for regulation of expression under varying supercoiling. Evidently, the detailed structural comparison between RIG and RRG promoters is needed to elucidate the mechanisms of their action. The extensive analysis by Peter et al. [8] provides the opportunity to assess the relationship of features observed in the corresponding promoter sequences with expression.

Our approach was based on several previous observations. Trifonov and Sussman [12] proved that the distribution of particular dinucleotides in genomic sequences is phased with the B-DNA helix pitch and that the corresponding helical periodicity is nearly universal in genomic sequences of all organisms. AT tracts phased with the B-helix pitch have been related to bending of double-stranded DNA. Since then, the effect of bent DNA on transcription regulation has been extensively studied [13–16]. It has also been shown that the characteristic period of helical periodicity in DNA sequences depends on the sign of supercoiling. Supercoiling is negative in Bacteria, whereas in hyperthermophilic Archaea, reverse gyrase induces positive supercoiling of the chromosome. The negative supercoiling shifts the double-helix pitch from ~10.3–10.5 bp (free B-form DNA) to ~11 bp, whereas the positive supercoiling shifts it toward ~10 bp. The shifts in the double-helix pitch are assumed to be reflected in the corresponding periodicities of DNA sequences. Such changes in periodicities of DNA sequences have been actually identified on the genome-wide scale [17–20]. Previously, we have shown that stretches with ~11 bp periodicity and ~10 bp periodicity interleave in the genomic sequence of *E. coli* and that the net bias to ~11 bp is attained by the abundance of ~11-bp stretches [21]. It has also proved that these periodicities were strongly pronounced in a part of SSG promoters. In this paper we present a systematic comparison of structural features in RIG and RRG promoters and study their correlations with the expression level.

**Results**

*Layout*

To make easier the presentation of the results, we describe briefly the main content and the relationships between topics below. First, significant features in the sequences of RIG and RRG promoters should be identified and mutually compared. Specifically, the hidden periodic patterns in DNA sequences modified by mutations and insertions/deletions may affect the binding of RNA polymerase and transcription factors with promoter DNA and thus affect the transcription. The surrounding regions (in particular, region upstream of promoters) also participate in transcription regulation. Therefore, the analysis of periodicities in the vicinity of promoters is informative and should be extended to these regions as well. The characteristic features found in the spectra for RIG promoters are expected to be related to the enhancement of expression under relaxation of chromosome supercoiling, whereas that of found for RRG promoters are expected to be related to the decrease of expression. This suggestion proved to be hold in the case of helical periodicities. Despite the small shift in helical periodicities for RIG and RRG promoters, the character of correlations changes inversely. The difference in correlations contains information on molecular mechanisms of SSG expression (Discussion). We developed the special scheme for analysis of correlations between the expression levels and the intensities of periodicities in promoter sequences, which is described in the third topic of this section. The two concluding topics are aimed at the detailing of the found features. In one of concluding topics we compare the helical periodicities for RIG and RRG promoters associated with different sigma factors of RNA polymerase, whereas the following topic is devoted to the analysis of helical periodicities with refined resolution.

*Periodicities in promoter sequences of supercoiling-sensitive genes*

In RegulonDB [22] we found 43 promoters for 26 RIG and 107 promoters for 86 RRG. The corresponding promoters were picked up from RegulonDB according to the annotation of SSG genes with statistically significant variation in expression level listed in supplemental materials to Ref. [8]. Search was performed by the gene notation. The list of promoters and related information are presented in Supplemental file 1 to this paper. Both sets are representative enough and allow us to juxtapose the features inherent to two groups of promoters. The analysis was performed with Fourier transform of DNA sequences (Materials and Methods). We used the format of sequences similar to PromEC database [23], in which the promoters were associated with the region (−75, +25) from the transcription start (TS).

Fourier spectra averaged over the two promoter sets are shown in Fig. 1. The AT regularities were characterized by the sum of structure factors $f_{AA} + f_{TT}$ (Materials and Methods), whereas the sum $f_{CC} + f_{GG}$ represented the GC regularities. Both sums remain invariant with respect to replacement of direct and complementary DNA sequences. The high spectral peaks corresponding to harmonics with numbers $n = 1, 2$ (periods $p = 101, 50.5$ nt; these units will always be used for the periods and will tacitly be implied in what follows) (Materials and Methods, Eq. (1)) are associated with variations in nucleotide composition on the window scale. Such variations in AT content are typical of both groups of promoters and, generally, of all AT-rich promoters in the *E. coli* genome. The same concerns the variations in GC content for RIG promoters. In Refs. [21, 24] the significant periodicities $p = 20.2–14.4$ ($n = 5–7$) indicating the spacer-period patterns (determined by the distance between the canonical –35 and –10 elements of the promoter) have been observed in the complete set of *E. coli* promoters. In the studied sets of RIG and RRG promoters (which are the subsets of the complete set) these periodicities appeared to be insignificant. The high peaks at $n = 34$ ($p \approx 3.0$) correspond to the 3-nt periodicity nearly universal in the protein-coding regions [25, 26]. The most interesting feature is the difference in intensities of the most pronounced helical periodicities in AT spectra, $n = 10$ ($p = 10.1$) for promoters of RIG and $n = 9$ ($p = 11.2$) for those of RRG. Correlations between the intensities of periodicities $p = 10.1$ and $p = 11.2$ proved insignificant.

The subsequent analysis is mainly focused on AT spectra due to their relationship with bending of DNA helix and cooperative binding with transcription factors. The analysis of averaged AT spectra was extended to a ±300 nt vicinity of TS with step = 1 and sliding window = 101. The distance was measured from the position of the 5'-end of a 101-nt window relative to TS. The corresponding averaged spectra are shown in Figs. 2A and 2B. In RegulonDB we found 43 promoters for 26 RIG and 107 promoters for 86 RRG. To assess the statistical significance of the observed regularities, the obtained values should be compared with the expected characteristics for the corresponding sets of random sequences of the same nucleotide composition. Standard deviation for sum $f_{AA} + f_{TT}$ averaged over the set of $P$ random sequences is equal to $(2/P)^{1/2}$ (Materials and Methods). Taking three standard deviations as an approximate analog of extreme value statistics, the heights exceeding the mean plus three standard deviations, or respectively 2.648 (Fig. 2A) and 2.411 (Fig. 2B), can be considered significant. Nearly all significant periodicities grow stronger upstream of TS, except for the coding periodicity $p \approx 3.0$ ($n = 34$), growing stronger in the region downstream of TS. These trends are in line with the similar observations for the complete set of promoters [16, 21]. Fig. 2C characterizes the difference between spectra for RIG and RRG promoters. The divergence between the mean intensities of periodicities $p = 10.1$ and 11.2 for RIG and RRG promoters is presented separately in Fig. 2D and clearly reveals the opposite signs of the difference within the region from $-150$ to $-50$ relative to TS. The corresponding dependencies for AT content and spectral entropy (Materials and Methods) are presented in Supplemental file 2.

*Correlations between expression log-ratios and periodicities in promoter sequences*

The periodic patterns in promoter sequences affect the binding of RNA polymerase and transcription factors with promoter DNA and thus may affect the expression level. The experimental data by Peter et al. [8] provide an opportunity to relate the expression level with the features observed in nucleotide sequences in the vicinity of the promoters. The relevant quantitative analysis is non-trivial: (i) the expression data are noisy; (ii) the expression varies (generally, non-monotonously) with time, temperature, or concentration of agents; and (iii) the genes may be transcribed from several different promoters, and it is unknown which promoter is active in particular conditions. Therefore, the search of potential correlations requires application of statistical methods, expert assessment, and extensive cross-checking. In this section we reproduce the summary of such bulky analysis.

In the experiments by Peter et al. [8] chromosome relaxation was attained by the action of antibiotics norfloxacin or novobiocin inhibiting gyrase and topo IV. Also they used a temperature-sensitive strain in which gyrase was inhibited at 42°C. The runs of four different experiments included: (i) gene expression at times $t = 2, 5, 10$, and 20 min after temperature shift in the temperature-sensitive mutant (Experiment 1); (ii) gene expression measured after addition of 15 μg/ml norfloxacin to an isogenic wild-type strain at times $t$ = 2, 3, 4, 5, 7, 10, 15, 20, and 30 min (Experiment 2); (iii) gene expression at times $t$ = 2, 5, 10, 20 or 30min after addition of 50 μg/ml norfloxacin to an isogenic wild-type strain (Experiment 3); and (iv) gene expression at fixed time $t$ = 5 min and varying concentrations of novobiocin (Novo) = 20, 50 or 200 μg/ml in the wild-type strain (Experiment 4). As a measure of impact of chromosome relaxation on expression, we chose the maximum expression log-ratio in each run for RIG and the minimum expression log-ratio for RRG. For the convenience of the reader, the corresponding expression log-ratios from Ref. [8] used in our analysis are reproduced in Supplemental file 1. In subsequent correlational analysis, we used Spearman rank correlation coefficients, which provide a more robust statistical measure than Pearson coefficients. For the each of four experiments there are sets of maximum (RIG) and minimum (RRG) expression log-ratios. The Spearman correlation coefficients between maximum log-ratios for RIG in different experiments were rather weak and ranged within 0.15–0.40. The same held for the correlations between minimum log-ratios for RRG in different experiments. For the cross-check, we treated the different experiments as independent. First, we describe the scheme of correlational analysis for the genes with multiple promoters. Then, the dependence of correlations in the vicinity of promoters will be analyzed. The maxima or minima of correlations localized in the promoter region provide additional arguments in favor of the origin of correlations from promoters.

According to RegulonDB, 14 genes of the RIG set are transcribed from 1 promoter, 9 genes from 2 promoters, 1 gene from 3 promoters, and 2 genes from 4 promoters, whereas in the RRG set, 70 genes are transcribed from 1 promoter, 12 genes from 2 promoters, 3 genes from 3 promoters, and 1 gene from 4 promoters. There is no information about which of the promoters worked in a specific experiment. Moreover, it is possible that different promoters might be (in)active in different experiments, because for some experiments all the expression log-ratios for genes attributed to RIG turned out to be negative, and vice versa for RRG. The choice of one-to-one correspondence between transcribed gene and active promoter provides $2^9 \times 3 \times 4^2 = 24{,}576$ different combinations of promoters for RIG and $2^{12} \times 3^3 \times 4 = 442,368$ combinations for RRG. To study the correlations between the structural features in DNA sequences and the expression log-ratios, we considered all possible combinations for the each set of promoters. The table of resulting distributions of Spearman correlation coefficients for all different combinations of promoters in the each set is shown in Fig. 3. The correlations were studied at a slightly shifted position (–70) relative to the definition of promoter set (–75). The significance of correlations was assessed by comparing the median in obtained distributions with the threshold values corresponding to 5% significance of correlations for RIG and RRG sets, ±0.389 and ±0.212. To verify the criterion, we used the numerical simulations. In simulations each promoter sequence was replaced by the random sequence of the same nucleotide composition. Then, the correlational analysis was performed for the random sequences along the lines as above. This criterion yielded significant anti-correlations (or correlations of negative sign) between the maximum expression log-ratios for RIG and the spectral entropy (Experiments 3 and 4), indicating the presence of signal repeats in RIG promoters related to the increase of expression. For RRG promoters, the significant anti-correlations were observed between minimum log-ratios and the intensity of periodicity $p = 11.2$ (Experiment 4, Average) and spectral entropy (Experiment 3, Average).

The analysis of correlations was extended to the vicinity of promoters and to the harmonics in the range of numbers $n = 1$–$11$. Furthermore, we studied the correlations between expression and AT content as well as the correlations between expression and spectral entropy. The set of promoters nearest to the translation start was chosen as a reference. The correlations for this set were about the median lines for distributions in Fig. 3. In this sense, the chosen set may be considered a typical representative. The curves in Fig. 4 show the correlations averaged over four experiments. The standard deviation for Spearman random correlations is $1/\sqrt{P-1}$, where $P$ is the number of promoter sequences in the set. If averaging is additionally performed over $N$ experiments, the corresponding deviation should be multiplied by the factor $1/\sqrt{N}$. The scales in Fig. 4 correspond to the range of two standard deviations, which is different for RIG and RRG sets. Attention should be paid to the significant (exceeding the threshold of two standard deviations) rises and falls of correlations localized in the vicinity of promoters (position −75). The reproducibility of observed correlations for all experiments enhances their statistical significance. The typical examples of reproducible and non-reproducible dependencies are shown in Supplemental file 3C. For RIG promoters we found significant reproducible correlations between expression log-ratios and AT content and similar weaker correlations for periodicity $p = 101$ ($n = 1$). The most striking effect was observed for periodicity $p = 50.5$ ($n = 2$). These anti-correlations were strongly peaked in the vicinity of promoters and attained a minimum down to −0.7 within the range from −100 to −75 relative to TS (Experiments 2 and 3; Supplemental file 3C). The anti-correlations for $n = 2$ behaved similarly in all four experiments for RIG (Supplemental file 3C), whereas for RRG the corresponding much weaker minima were observed only for Experiments 2 and 3. The study of correlations in the vicinity of RIG promoters revealed also significant reproducible correlations with intensity of periodicity $p = 9.2$ ($n = 11$) in the upstream region (left from −75) (Supplemental file 3C). For RRG set the most salient feature referred to the periodicity $p = 11.2$, which revealed significant reproducible anti-correlations with expression of RRG peaked in the vicinity of promoters (Supplemental file 3C). The most interesting averaged dependencies for correlations are collected in Fig. 4. The complete information may be found in Supplemental file 3.

*Helical periodicities in promoter sequences and sigma factors of RNA polymerase*

The promoters are recognized by the sigma subunit of RNA polymerase. The sigma subunits of seven different types are known for *E. coli* [27–29]. The transcription of genes in the growth phase is normally associated with high levels of negative supercoiling and for most of the genes is driven by $\sigma^{70}$, whereas relaxation of the chromosome is correlated with enhanced usage of $\sigma^{38}$ [10, 11, 30]. The binding of both factors with the promoter proved to be sensitive to the spacer between the −35 and −10 elements [31, 32]. We compared the helical periodicities in promoters recognized by different sigma factors as well as the percentage of sigma factors associated with RIG and RRG promoters (Fig. 5). For the complete dataset in RegulonDB, AT periodicity $p = 11.2$ dominates in promoters recognized by $\sigma^{70}$, whereas in promoters recognized by $\sigma^{38}$ the intensity of periodicity $p = 10.1$ appears to be higher in accordance with topological predictions (Fig. 5A). Nevertheless, the hypothesis that these periodicities will dominate in the corresponding subsets independent of belonging to RIG or RRG fails (Figs. 5B and 5C). As expected, the percentage of promoters recognized by $\sigma^{70}$ for RRG exceeds that for RIG, while the percentage of promoters recognized by $\sigma^{38}$ is lower for RRG relative to RIG (Fig. 5D).

*Helical periodicities in the range of periods 9.5−11.5 nt*

The window length for Fourier analysis was chosen to be 101 bp for the reasons explained previously [21]: the window about 100 nt is perhaps the shortest window resolving generic periodicities and ensuring the locality of analysis; this window is a bit longer than the RNA polymerase contact region and is slightly shorter than the persistence length of dsDNA (~150 bp); it agrees with the conventional definition of promoter region [23]. The corresponding periods $p$ are obtained through the length of a sequence $L$ and harmonic number $n$ by $p = L/n$ (Materials and Methods). Therefore, Fourier transform with 101-nt window resolves only the periods 10.1 and 11.2 for the harmonics $n = 10$ and 9. The experimentally determined pitch of free B-form DNA in solution is 10.3–10.5 [33] and falls in the intermediate range. Choosing the harmonic number $n = 10$ and varying length in the range 95-115 with step 1 allows one to resolve the periods in the range 9.5–11.5 with step 0.1.

The dependence of mean intensity on period (resonance plot) shows maxima at 10.1 for RIG promoters and at 11.3 for RRG promoters, with very flat tailing into longer periods for the RRG set (Supplemental file 4). The window of length 101 nt is optimal from the viewpoint of resonance plots for the promoter set. All SSG promoters with statistically significant helical periodicities are listed in Supplemental file 4. The refined analysis supports the conclusion that the helical periodicities in RIG promoters are shortened toward 10 nt, whereas those in RRG promoters are extended toward 11 nt. These trends are retained in the vicinity of promoters (Supplemental file 4).

If periodicities close to 10 and 11 nt in promoters affect the binding with transcription factors and RNA polymerase depending on DNA supercoiling, the correlations between the intensities of these periodicities and the expression log-ratios under relaxation of supercoiling are expected to be positive for $p = 10.1$ and negative for $p = 11.2$ in both groups of promoters. Similar trends may be seen, though not robustly and insignificantly, for periodicity $p = 10.1$ in both groups (Figs. 3 and 4) and more distinctly for periodicity $p = 11.2$ in the case of RRG promoters. The verification of hypothesis that the change in periodicity from $p = 10.1$ to 11.2 leads to the reversal of correlations with the expression may be improved by uniting the sets of RIG and RRG promoters for better statistics. For the united set, we found significant reproducible correlations between the intensity of periodicity $p = 10.1$ and the expression log-ratios, peaking in the vicinity of promoters, whereas for periodicity $p = 11.2$ significant reproducible anti-correlations were observed, showing minimum in the vicinity of promoters (Figs. 6A and 6B; Supplemental file 4). The study of correlations between the expression log-ratios and the intensity of helical periodicities in the range 9.5–11.5 also revealed the strongest effects at ~10 and ~11 nt in the vicinity of promoters (Fig. 6C).

## Discussion

Our study proves the clear bias of periods for underlying helical periodicities in DNA sequences of RIG and RRG promoters to 10 and 11 nt, respectively. The abundance of promoters with dominating periodicity ~11 nt in the complete set of RegulonDB (see Ref. [21]) correlates with that of RRG [8] and, generally, with the important role of negative supercoiling during transcription in *E. coli*. The difference in the intensities of periodicities close to 10 and 11 nt persists in the two groups of promoters almost independently of sigma factor types (Figs. 5B and 5C). The prevalence of periodicity close to 11 nt in the complete set of promoters recognized by $\sigma^{70}$ (Fig. 5A) may be related to the abundance of RRG promoters in this subset, with a reverse situation for promoters recognized by $\sigma^{38}$. The data in Figs. 5B and 5C raise the following question. May different periodicities be recognized by sigma factors of the same type, and how can this be reconciled with the sensitivity of sigma factor to DNA sequence between the −35 and −10 elements? The possible explanation may be related to factors that bind RNA polymerase rather than DNA [34] and switch the binding to different DNA motifs.

Torsional tension induced by negative supercoiling tends to unwind DNA helix and facilitates transcription initiation. Negative tension extends also the pitch of DNA helix toward ~11 nt. Such periodicities appear to be typical of RRG promoter sequences. The processive motion of RNA polymerase induces the positive supercoiling downstream and negative supercoiling upstream of the transcribed DNA stretch [35–37]. The corresponding positive and negative supercoiling is relaxed by gyrase and topo I, respectively. Additionally, topo IV decatenates tangled duplicated chromosomes and is involved in removal of positive supercoils. In the chromosome of wt *E. coli* cells negative supercoils do not propagate to regions more distant than 0.8 kb, whereas the modifications produced by positive supercoils can be detected up to 4 kb away in the chromosome [38]. If gyrase and topo IV are inhibited by antibiotics, as in the experiments by Peter et al. [8], the positive supercoiling cannot be relaxed and its level downstream of efficiently transcribed genes may be high enough. The tension induced by positive supercoiling shortens the pitch of DNA helix toward ~10 nt. Such periodicities prove to be typical of RIG promoter sequences.

The genetic consequences of the observed changes in the periodicities of SSG promoters may be illustrated by promoters for genes *gyrB* and *topA* coding for subunits of gyrase and topo I, respectively. The *gyrB*P promoter has strong periodicity ~10 nt, whereas the *topA*P3 promoter has strong periodicity ~11 nt. Relaxation of the chromosome upregulates *gyrB* and downregulates *topA* as a feedback control [5–7]. The periodicity in *topA*P3 is close to the pitch of the DNA double helix under negative supercoiling, whereas the periodicity in *gyrB*P tends to the pitch of the double helix under positive supercoiling and is rather close to the lower bound of the pitch in relaxed DNA, 10.3 bp. These examples may be considered as a paradigm for the relationships between periodicities in RRG and RIG promoters, supercoiling, and expression. We also observed a strong periodicity ~11 nt in the sequence of promoter for the *fis* gene coding for architectural protein FIS in *E. coli*. The expression of *fis* is maximal at high levels of negative supercoiling [39]. Additional examples can be found in Supplemental file 4.

Supercoiling can also be produced by sequence-specific DNA-binding proteins [40]. A part of periodicities in Supplemental file 4 may reflect cooperative interaction with DNA-binding proteins. Specifically, toroidal supercoiling produced by wrapping DNA around a protein [9] may also induce positive superturns in DNA double helix.

The periodicities close to 10 and 11 nt proved to be typical of the complete sets of promoters recognized respectively by sigma factors $\sigma^H$ and $\sigma^A$ of RNA polymerase in *Bacillus subtilis* [41]. The factor $\sigma^A$ corresponds to $\sigma^{70}$ in *E. coli* and the correspondence between periodicities for this factor appears to be the same as that shown in Fig. 5A for $\sigma^{70}$. The factor $\sigma^H$ participates in the expression of genes at early sporulation, and there is no counterpart for it in *E. coli*. For this factor, periodicity ~10 nt proved to be typical. It may be expected that such selection in periodicities should be ubiquitous for SSG promoters in a wide class of species. These observations indicate the potential role of SSG in antibiotic-resistant bacteria, because many antibiotics inhibit bacterial gyrase and topo IV, which is in the direct correspondence with the methods used in Ref. [8] for chromosome relaxation.

We found reproducible positive correlations between expression enhancement for RIG genes and AT content in the promoter region as well as a bit lower correlations with the intensity of the longest periodicity $p = 101$ ($n = 1$), whereas the correlations with intensity of periodicity $p$

= 50.5 ($n$ = 2) turned out to be reproducibly and distinctly negative, with sharp minima in the vicinity of promoters (Results; Fig. 4; Supplemental file 3). The anti-correlations with intensity of periodicity $p$ = 50.5 ($n$ = 2) were the strongest effect that we found in our correlational analysis. Bacterial RNA polymerase contacts DNA by the σ factor recognizing the spacer between the −35 and −10 elements and by the C-terminal domains of the α subunits recognizing the UP element positioned between −60 and −40 [26, 40]. The total length from −60 to −10 coincides with period $p$ = 50.5 and may refer to the impact of this periodicity on the expression. The corresponding correlations with large-scale AT variations in promoters of RRG were pronounced much less robustly and significantly.

The significant anti-correlations between spectral entropy and expression for RIG (Fig. 3A, the third line, Experiments 3 and 4) reveal the connection between the periodic patterns in promoter sequences and the stronger enhancement of expression. The similar relationship between periodic patterns and the more suppressed expression of RRG would lead to the positive correlations and is (insignificantly) fulfilled only for Experiment 4 (Fig. 3B, the third line). The different periodicities may be responsible for either correlations or anti-correlations with expression (Results; Figs. 3 and 4; Supplemental file 2). As the spectral entropy presents the integral measure of all periodicities (Materials and Methods), its resulting correlations with expression depend on the relative contribution of positive and negative correlations with particular periodicities. The advantage of approach based on the spectral entropy consists in synergetic assessment of impact produced by different patterns.

Our study was aimed primarily at identifying the structural features inherent to DNA sequences of promoters for the genes with expression enhanced and decreased under relaxation of chromosome supercoiling in *E. coli*. At the next step, this technique can be applied to the search for SSG and to the discriminant analysis allowing differentiation between RIG and RRG promoters. The developed scheme of correlational analysis is universal and may be applied to data mining in gene expression analysis with expression microarrays. Taking into account the huge data on DNA sequences and expression levels stored in databanks and scattered throughout numerous publications, we hope that our study may initiate the further regular investigations on the relationship between structural features in DNA sequences and gene expression levels.

## Materials and Methods

*E. coli* K-12 promoter sequences, transcription and translation start sites, and information about sigma factor(s) recognizing the given promoter were retrieved from RegulonDB release 7.0 [22]. The genomic sequence of *E. coli* K-12 MG1655 was retrieved from GenBank (ftp://ftp.ncbi.nih.gov/genbank/genomes). The list of SSG in *E. coli* and their expression log-ratios were taken from Additional files to the publication by Peter et al. [8]. Spearman correlation coefficients and their box-and-whiskers distributions were obtained using modules Statistics::RankCorrelation 0.1203, Statistics::Basic 1.6607, and GD::Graph::boxplot from Comprehensive Perl Archive Network (http://www.cpan.org/). Fourier transform of DNA sequences and the corresponding bioinformatics analysis were performed with programs developed by ourselves.

*Fourier transform of DNA sequences*

The periodic patterns in promoter sequences were studied by Fourier transform following Refs. [21, 43, 44]. Fourier harmonics corresponding to nucleotides of type $\alpha$ ($\alpha$ is A, C, G, or T) in a sequence of length $L$ are calculated as

$$\rho_\alpha(q_n) = L^{-1/2} \sum_{m=1}^{M} \rho_{m,\alpha} e^{-iq_n m}, \quad q_n = 2\pi n/L, \quad n = 0,1,...,L-1 \qquad (1)$$

Here $\rho_{m,\alpha}$ indicates the position occupied by the nucleotide of type $\alpha$; $\rho_{m,\alpha} = 1$ if the nucleotide of type $\alpha$ occupies the *m*-th site and 0 otherwise. The nucleotides modified by methylation, hydroxylation and/or glycosylation should be considered separately and denoted by the special letters. The extension of Fourier analysis to the sequences with modified nucleotides is straightforward, but is beyond our aims in this paper. The amplitudes of Fourier harmonics (or structure factors) are expressed as

$$F_{\alpha\alpha}(q_n) = \rho_\alpha(q_n)\rho_\alpha^*(q_n) \tag{2}$$

where the asterisk denotes complex conjugation. The zeroth harmonics, depending only on the nucleotide composition, do not contain structural information and will be discarded below. The structure factors will always be normalized with respect to the mean spectral values, which are determined by the exact sum rules,

$$f_{\alpha\alpha}(q_n) = F_{\alpha\alpha}(q_n)/\overline{F}_{\alpha\alpha}; \quad \overline{F}_{\alpha\alpha} = N_\alpha(L - N_\alpha)/L(L-1) \tag{3}$$

where $N_\alpha$ is the total number of nucleotides of type $\alpha$ in a sequence of length $L$. The spectrum of structure factors (2) is symmetrical relative to $q_n = \pi$,

$$f_{\alpha\alpha}(q_n) = f_{\alpha\alpha}(2\pi - q_n) \tag{4}$$

Therefore, the spectrum can be restricted to the left half, $q_n \leq \pi$ or $1 \leq n \leq L/2$. In the main text the spectra should be understood in this sense. The characteristic period and the harmonic number are related as $p = L/n$. Generally, the significant periodicities should be identified not only by the singular high peaks in Fourier spectra but also by the sets of equidistant harmonics with the numbers $n, 2n, ..., rn \leq L/2$ [43–47]. The latter method needs, however, rather tedious expert assessment of the possible contributions of the significant shorter periodicities into the sums. In this work, the sum of equidistant harmonics is used formally as a cross-check.

Commonly, the periodic patterns in DNA sequences are strongly randomized by point mutations and insertions/deletions during molecular evolution. The typical scenario may be as follows: (i) take a stretch of any tandem repeats, e.g., ATG|ATG|ATG...; (ii) replace randomly a part of nucleotides in this stretch (generally, random insertions/deletions should also be taken

into account). The resulting stretch can be defined as a sequence with hidden periodicities. The frequency of random replacements may be biased in the different sites of a repeat. For instance, the random replacements in the second and third positions in the example above would yield the patterns ANN|ANN|ANN..., where N is any nucleotide. Fourier transform provides useful tool of the search for hidden periodicities. Correct judgment on the significance of hidden periodicities revealed by Fourier analysis needs application of proper statistical criteria. Random sequences of the same nucleotide composition serve as a reference for assessing the observed regularities in a given DNA sequence. Averaging over $P$ spectra for random sequences yields, in the limit of large $P$, a Gaussian distribution for each structure factor with the mean and standard deviation $<f>=1, \sigma(f)=1/\sqrt{P}$. The distributions for structure factors with different wave numbers may be considered independent. The pronounced peaks in the whole spectra should be compared with the singular outbursts in the spectra for random nucleotide sequences by extreme value statistics. Throughout the paper we use a 5% threshold of statistical significance. The normalized differences of harmonics in Figs. 2C and 2D were defined, respectively, as

$$\Delta \bar{f}_{norm}(q_n) = [(\bar{f}_{AA}(q_n) + \bar{f}_{TT}(q_n))_{RIG} - (\bar{f}_{AA}(q_n) + \bar{f}_{TT}(q_n))_{RRG}]/\sqrt{2}(1/P_{RIG} + 1/P_{RRG})^{1/2} \quad (5)$$

$$\Delta f_{norm} = [(\bar{f}_{AA}(q_{10}) + \bar{f}_{TT}(q_{10})) - (\bar{f}_{AA}(q_9) + \bar{f}_{TT}(q_9))]/2(1/P)^{1/2} \quad (6)$$

obeying approximately Gaussian statistics with zero mean and unit standard deviation for the corresponding random sequences. Here $P$ is the number of sequences in the corresponding promoter set and $\bar{f}$ means the harmonic averaged over set.

*Periodic patterns and spectral entropy*

The spectral entropy provides the quantitative measure of order/disorder in DNA sequence and is defined by the sum

$$S_\alpha = -\sum_{n=1}^{N} f_{\alpha\alpha}(q_n) \ln f_{\alpha\alpha}(q_n) \quad (7)$$

over spectrum (here $N$ is the integer part of $L/2$). Its values are strictly negative. The order in DNA sequence can be related to hidden periodic patterns. The lower (or more negative) values of spectral entropy indicate the higher ordering of DNA sequence or the more pronounced periodic patterns in comparison with random sequences of the same nucleotide composition. Otherwise, the higher (or more close to zero) values of spectral entropy indicate the higher frequency of point random mutations in the corresponding stretches or their stronger variability. The correlations with expression in the main text (Fig. 3) were studied for the sum $S_A + S_T$. The spectral entropy proved to be useful in assessment of the general difference between sequences for coding and non-coding stretches [48] or between genes and pseudogenes [49, 50]. The similar definition (7) holds for the spectra averaged over $P$ sequences.

## Acknowledgments

The authors are grateful to A.V. Galkin for stimulating discussions and for editing the text. This work was supported by the Molecular and Cellular Biology Program of the Presidium of the Russian Academy of Sciences.

## Appendix A. Supplementary data

Supplementary data associated with this article can be found, in the online version, at doi:

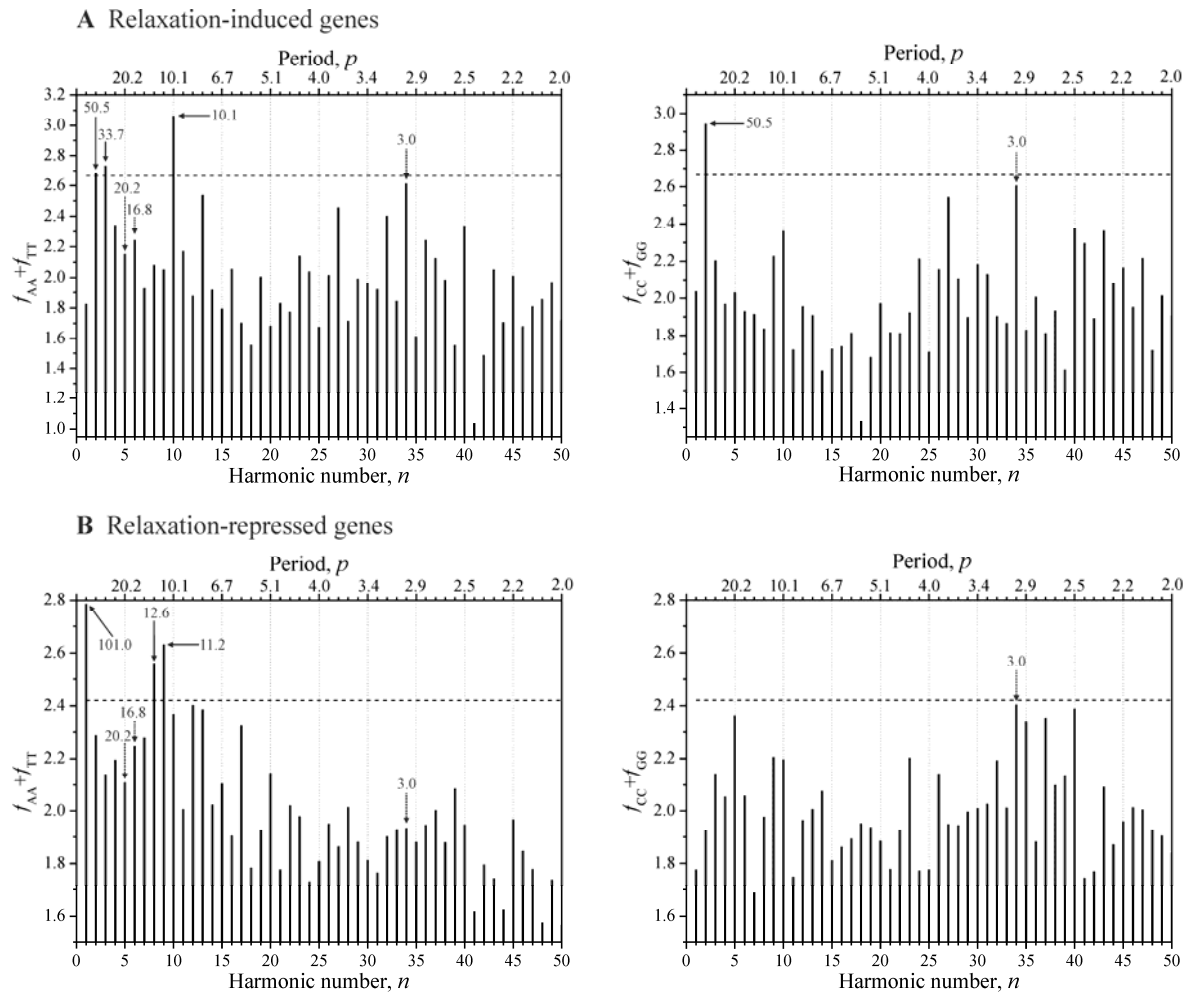

Fig. 1. Fourier spectra for AT and GC structure factors. The spectra were averaged over the sets of *E. coli* RIG and RRG promoters. The horizontal line corresponds to 5% probability that any harmonic in an averaged spectrum for random sequences of the same nucleotide composition as the counterpart promoters exceeds this level. The characteristic significant (solid arrows) and insignificant (broken arrows) periodicities are shown separately (see text). The shift in helical AT periodicities typical of RIG and RRG promoters from $p = 10.1$ to 11.2 indicates the difference in the supercoiling conditions for enhanced transcription of the corresponding genes.

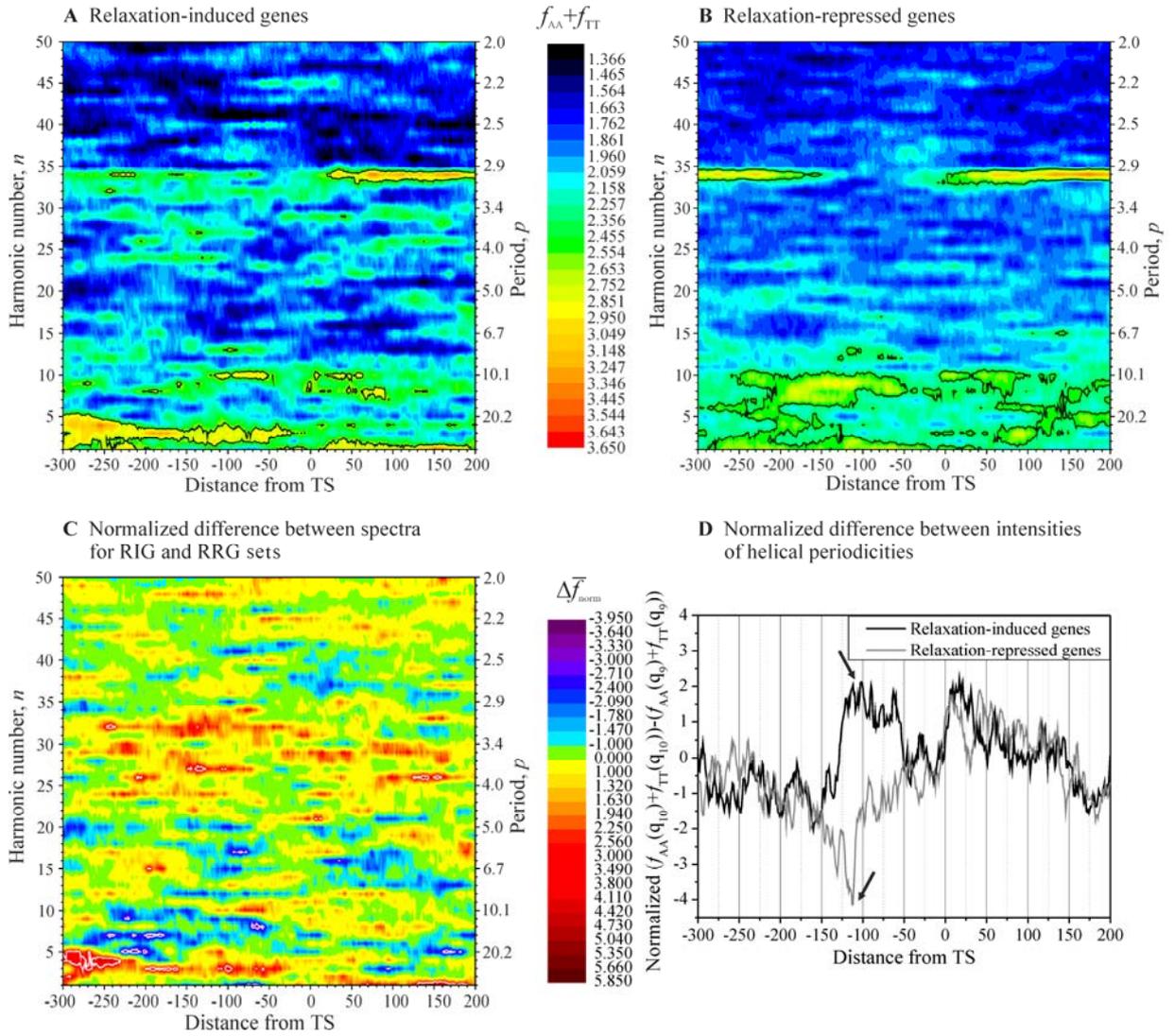

Fig. 2. Fourier spectra in the vicinity of promoters. (A and B) The averaged spectra for the sum of structure factors, $f_{AA}(q_n) + f_{TT}(q_n)$, characterizing intensities of different periodicities in the vicinity of promoters. The position is measured by the distance from the 5'-end of a 101-nt sliding window to the transcription start (TS). Averaging is performed over all sliding windows at given position from TS. (C) The normalized difference between mean intensities of the counterpart periodicities for RIG and RRG sets and (D) the normalized difference between mean intensities of harmonics corresponding to periodicities $p = 10.1$ ($n = 10$) and $p = 11.2$ ($n = 9$) in RIG and RRG sets (Eqs. (5) and (6), Materials and Methods). The black contours in panels A–B and the white contours in panel C mark statistically significant values. The arrows in panel D indicate the opposite divergence between the intensities of helical periodicities $p = 10.1$ and $11.2$ for RIG and RRG in the vicinity of promoter region (position $-75$), which can be related to the difference in the supercoiling conditions for enhanced transcription of the corresponding genes. Note the significant or close to them variations in the characteristic periodicities (marked in Fig. 1) for RIG and RRG promoters shown in panel C.

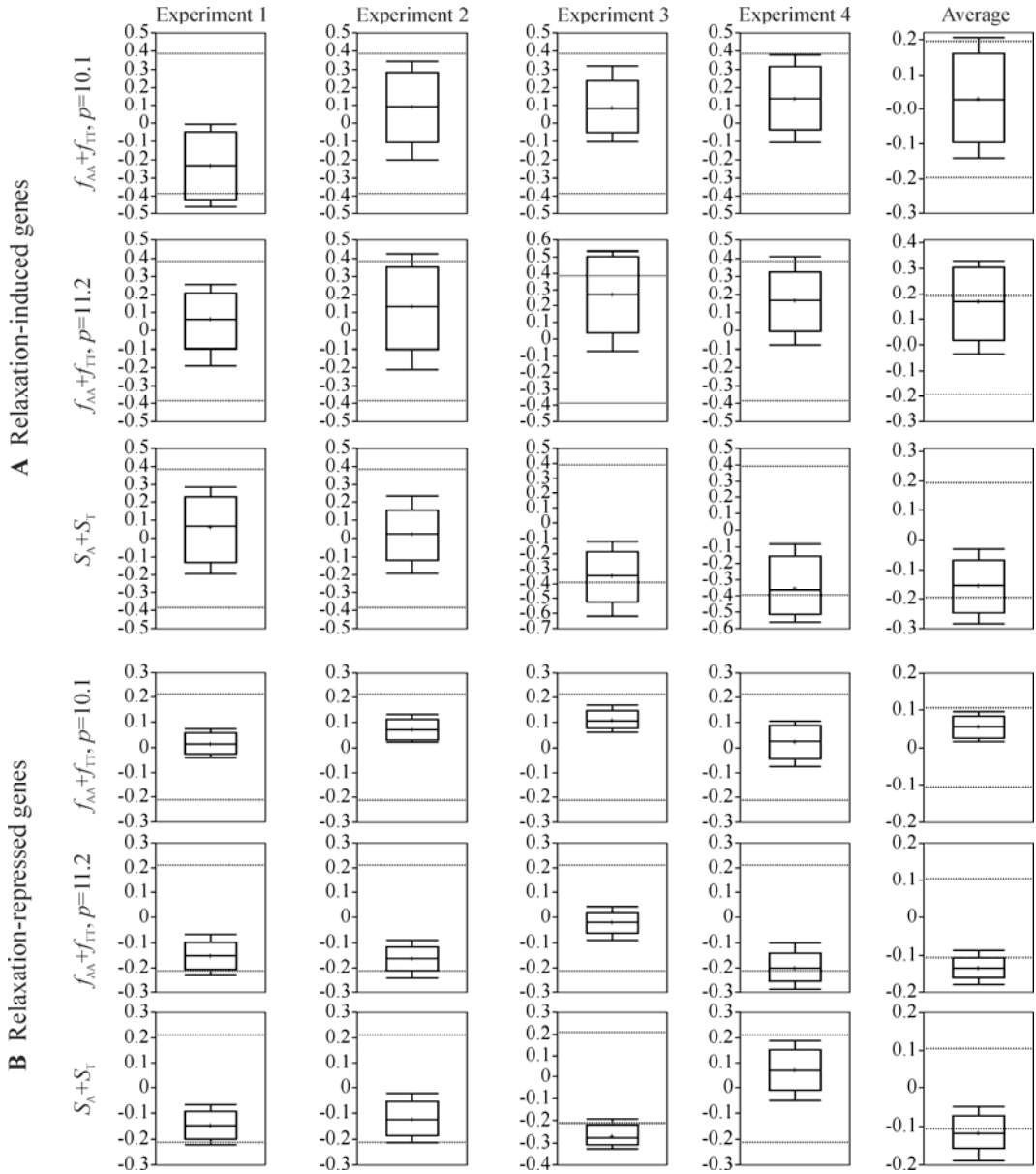

Fig. 3. Box-and-whiskers representation for distributions of Spearman rank correlation coefficients. The whiskers in each plot mark the min/max values of the distribution, the horizontal line denotes the median, and the cross denotes the mean. Boxes comprise 95% of data and should be compared with 5% significance thresholds, ±0.389 (RIG) and ±0.212 (RRG) (marked by the broken lines). The thresholds for the distributions averaged over four experiments are twice less. The correlations were calculated between intensities of periodicities $p$ = 10.1. 11.2 in promoter sequences and maximum (RIG) or minimum (RRG) expression log-ratios in each of four different experiments (Results). The correlations between spectral entropy (Materials and Methods, Eq. (7)) and expression are also presented in this figure (the third lines in the sets A and B). The data on the expression log-ratios were taken from Ref. [8]. Different experiments correspond to the different conditions of supercoiling relaxation (see text and [8]). The entire picture should be considered as an analog of two 3×5 tables of correlation coefficients. The positive correlations mean that the higher the intensity of a periodicity, the stronger the increase of expression under relaxation of chromosome supercoiling, whereas for the negative correlations (or anti-correlations) the relationship is reciprocal. The negative correlations with spectral entropy indicate the presence of signal repeats related to the increase of expression.

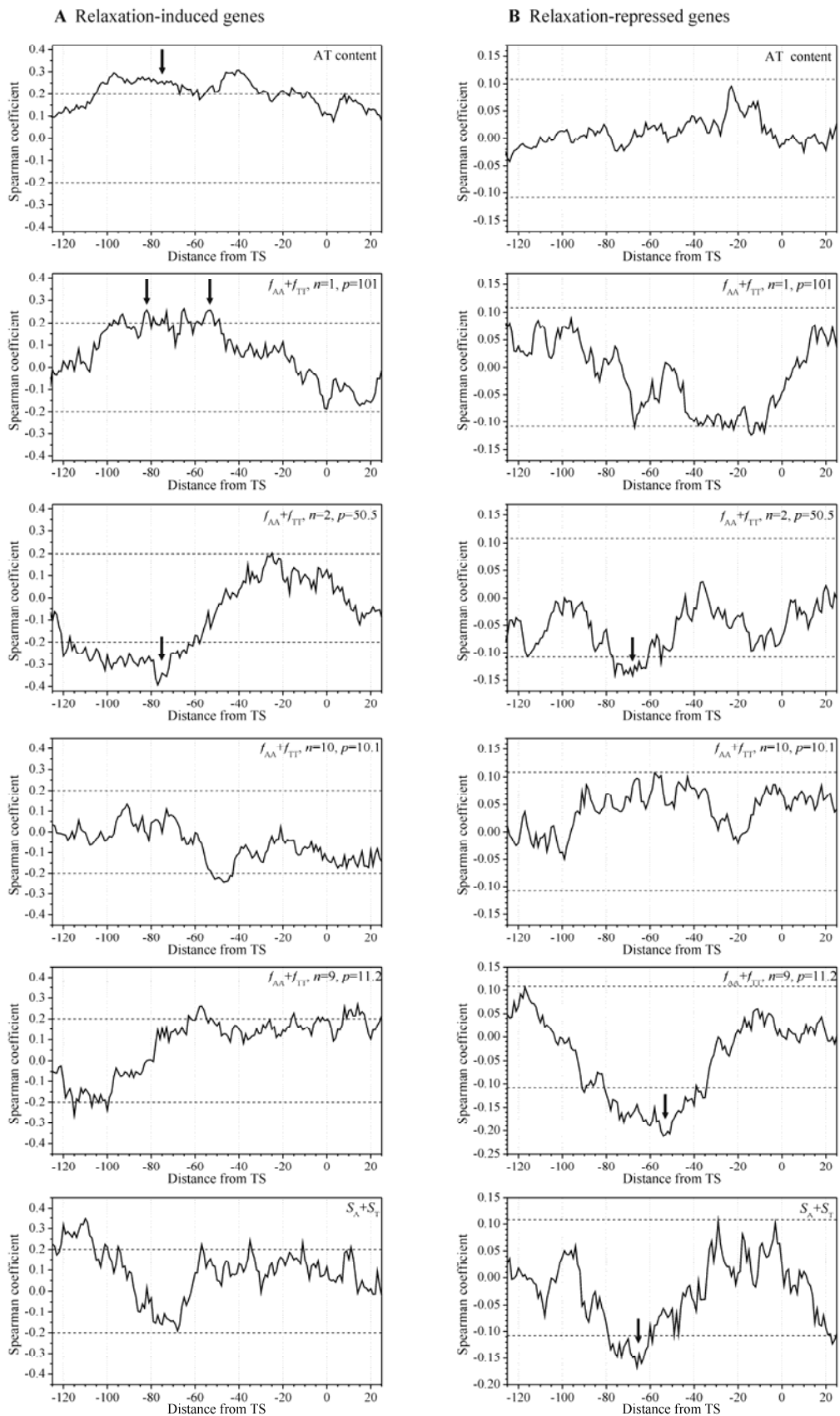

Fig. 4. Dependence of correlations on the distance from TS for the promoters nearest to translation start. The position is measured from the 5'-end of 101-nt window to TS. The correlations were calculated between AT content, intensities of periodicities, spectral entropy and maximum (RIG) or minimum (RRG) expression log-ratios in each of four different experiments [8]. The curves in this figure correspond to the correlations averaged over four

experiments. The horizontal broken lines mark two standard deviations for the random correlations and serve for the assessment of statistical significance. The positive correlations mean that the higher the intensity of a periodicity, the stronger the increase of expression under relaxation of chromosome supercoiling, whereas for the negative correlations (or anti-correlations) the relationship is reciprocal. The negative correlations with spectral entropy indicate the presence of signal repeats related to the increase of expression (bottom). The significant correlations reproducible throughout experiments and peaked in the vicinity of promoters (position –75) are marked by the arrows.

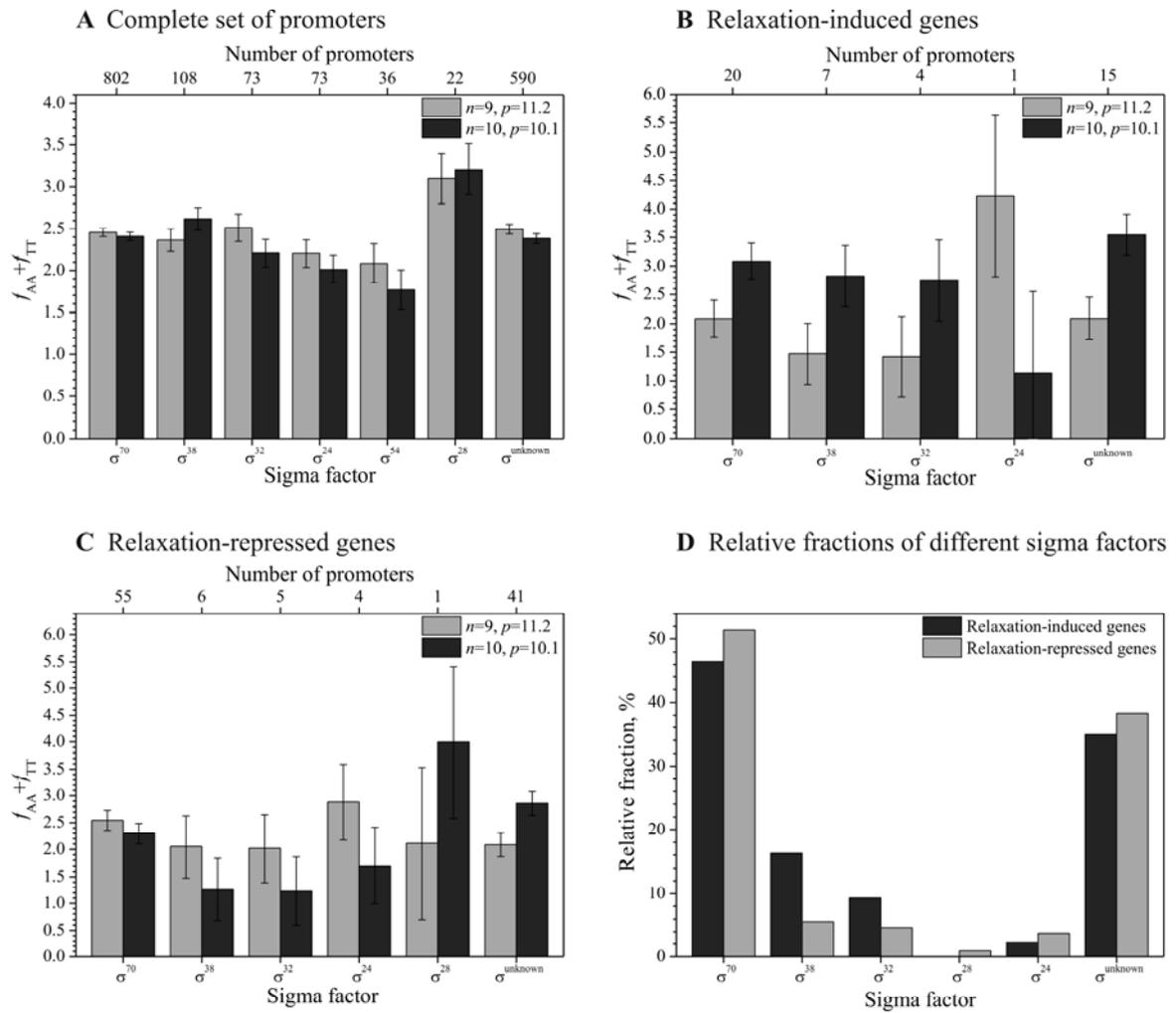

Fig. 5. Intensities of helical periodicities in promoters associated with different sigma factors. The mean intensities of harmonics $f_{AA}(q_n) + f_{TT}(q_n)$ corresponding to periodicities $p = 10.1$ ($n = 10$) and $p = 11.2$ ($n = 9$) in promoters associated with different sigma factors for the complete set of promoters in RegulonDB (A), for the promoters related to RIG (B), and for the promoters related to RRG (C). The flags on the bars refer to one standard deviation in the corresponding random sets. (D) The percentage of sigma factors associated with RIG and RRG promoters. The promoters associated with two or more sigma factors were counted in each subset. The results in the panels B–C show that the helical periodicities characteristic of RIG and RRG promoters persist also in the promoters associated with different sigma factors.

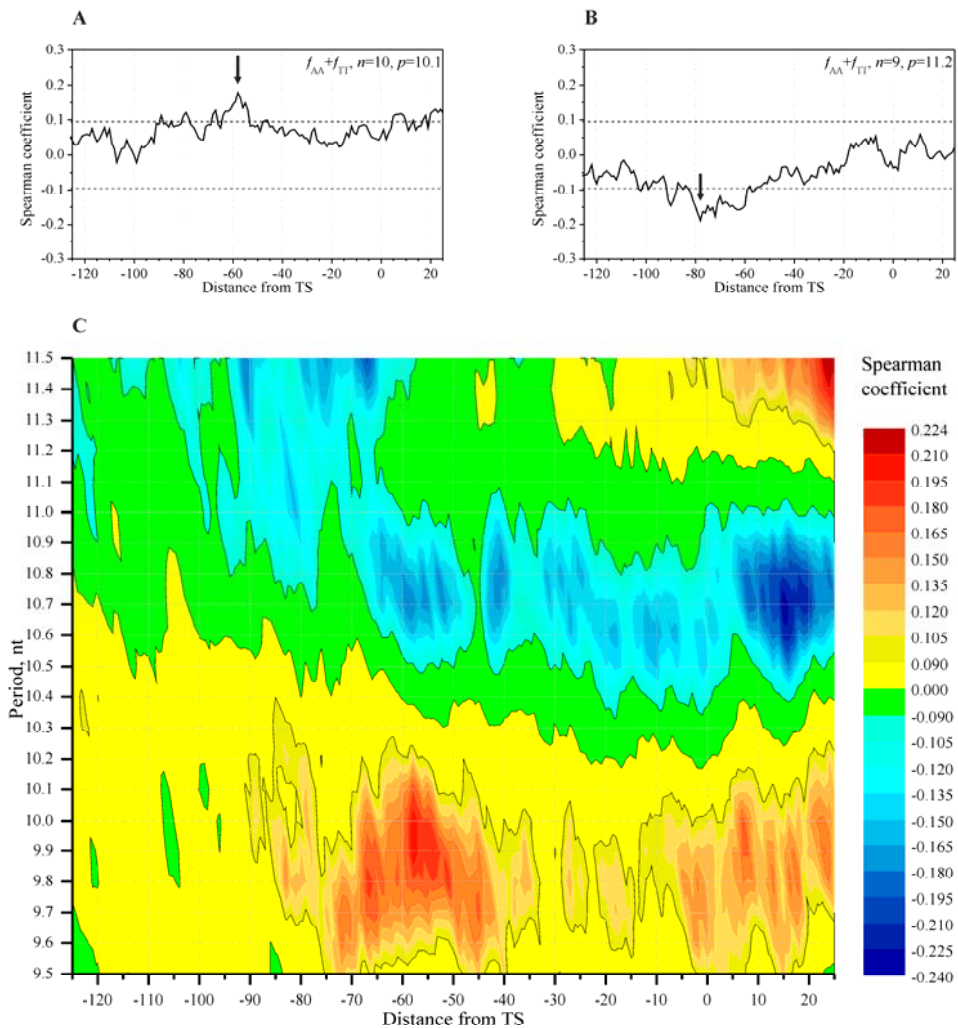

Fig. 6. Dependence of correlations between the expression log-ratios and the intensities of periodicities $p = 10.1$ (A) and $p = 11.2$ (B) on the distance from TS for the promoters nearest to translation start. The horizontal broken lines mark two standard deviations for the random correlations. The maximum (for $p = 10.1$) and minimum (for $p = 11.2$) of correlations (marked by arrows) indicate their origin from promoter region (position −75). (C) The general overview of correlations between the expression log-ratios and the intensities of helical periodicities within the range 9.5–11.5 in the vicinity of promoters. The contours denote the levels of correlations zero and plus/minus two standard deviations for the random counterparts. In all panels A–C the latter value serves for the assessment of statistical significance. The position is measured from the 5'-end of 101-nt window to TS. The RIG and RRG promoters were united into one set. The curves in all figures correspond to the correlations averaged over four experiments. The panels A–C show that the shift in helical periodicities from ~10 nt to ~11 nt leads to the inversion of correlations. If the helical periodicities ~10 nt correlate with the increase of expression under relaxation of chromosome supercoiling, the helical periodicities ~11 nt correlate with the decrease of expression.

# Supplemental files 2 & 3

## AT content in the vicinity of promoters

Average AT content in the vicinity of RIG and RRG promoters is presented below. The position is measured by the distance from the 5'-end of a 101-nt sliding window to TS. Averaging was performed over all sliding windows at given position from TS.

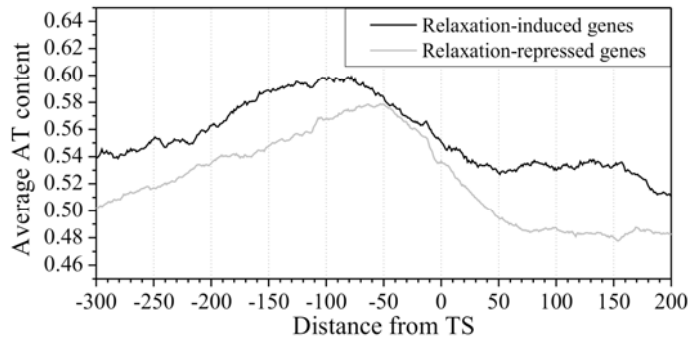

A higher AT content is inherent to promoters in *E. coli* and many other bacterial genomes. In the set of promoters for SSG, the maximal AT content for RIG promoters reached 0.599 at position −78, whereas the corresponding maximum for RRG promoters was 0.579 at −53. This difference in AT content is statistically insignificant. In the coding regions far from promoters, the AT content tends to the homogeneous value of 0.5 (compare the mean AT contents at the ends of the range shown in this figure). Note that the difference in divergence between mean intensities of periodicities $p = 10.1$ ($n = 10$) and $p = 11.2$ ($n = 9$) corresponding to RIG and RRG promoters was observed only in the vicinity of promoters within the range from −150 to −50 relative to TS but not within the coding regions, despite the comparable AT contents (see Fig. 2C and this figure). The dependence of AT content on position relative to the transcription start in the vicinity of promoters appears to be similar to that for the coding regions of SSG relative to the translation start [8].

**The normalized deviations of AT spectral entropy from the mean value corresponding to random sequences**

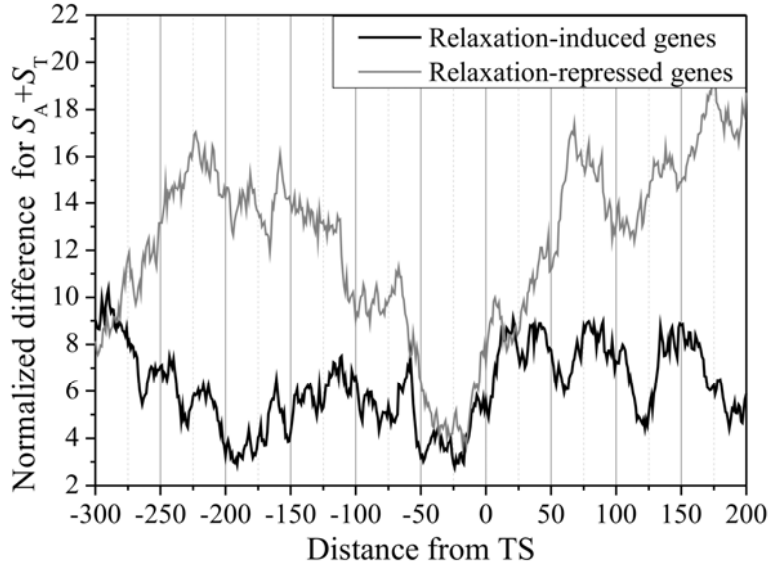

The normalized deviations of AT spectral entropy from the mean value corresponding to random sequences in the vicinity of promoters were calculated according to

$$z_S = (2<S_\alpha>_{random} - S_A - S_T)/\sqrt{2}\,\sigma(S_\alpha)$$

where spectral entropy related to the spectra averaged over $P$ promoter sequences $S_\alpha$ is defined by Eq. (7) in Materials and Methods. The corresponding mean value and standard deviation for averaged spectra of the random nucleotide sequences are $<S_\alpha>_{random} \approx -N/2P$; $\sigma(S_\alpha) \approx (N/2P^2)^{1/2}$ ($N$ is the integer part of the quotient [$L/2$] and $L$ is the length of a sequence). The normalized deviation obeys Gaussian statistics with zero mean and unit standard deviation for the random sequences. The deviations correspond to 101-nt sliding windows. The position is measured by the distance from the 5'-end of a 101-nt sliding window to the transcription start (TS). Spectral entropy related to the averaged spectra characterizes both the integral intensity of hidden periodicities in the promoter sequences and its variability over promoter set.

**A** Relaxation-induced genes

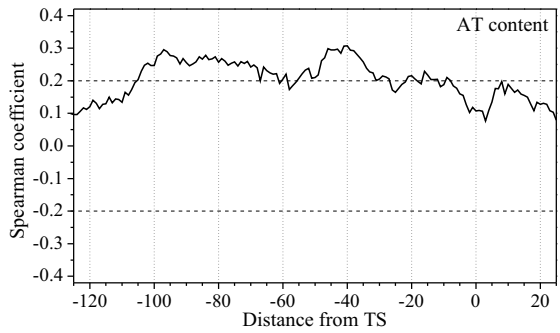
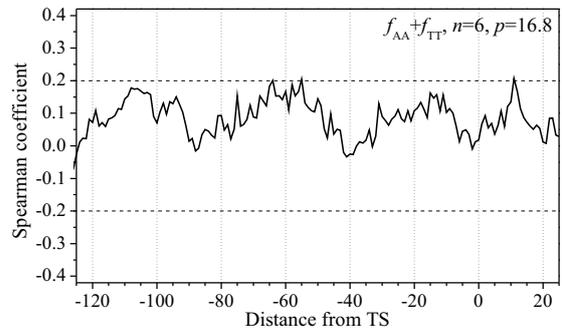
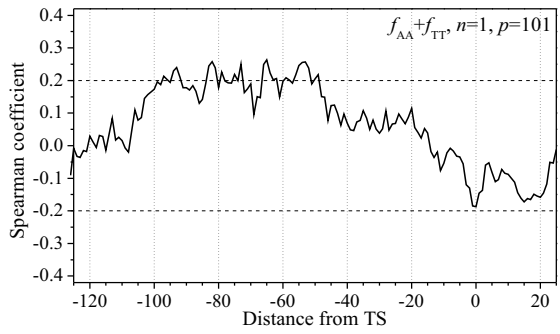
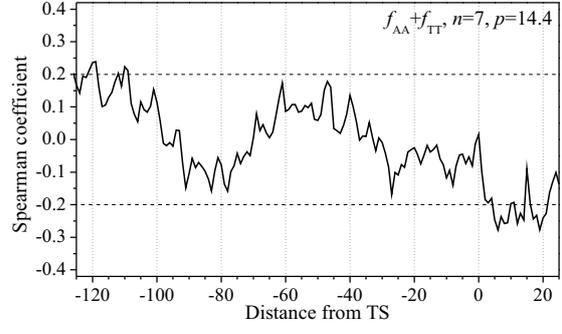
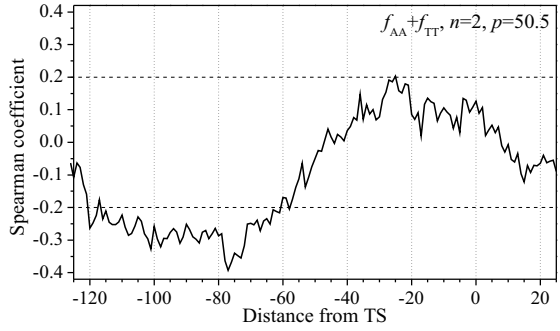
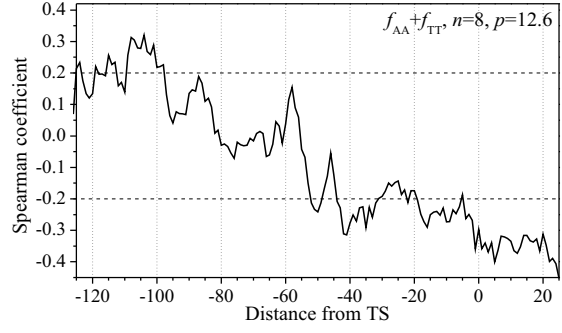
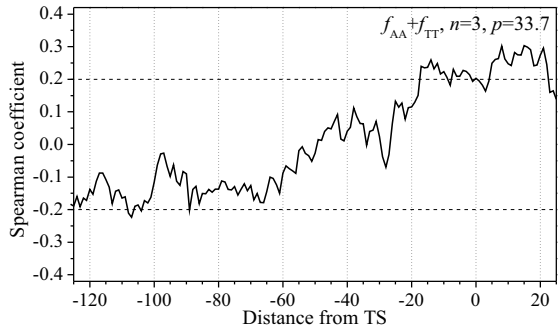
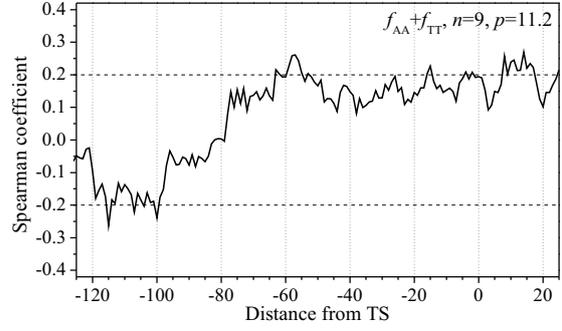
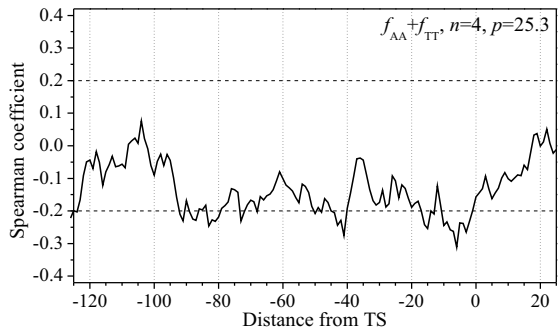
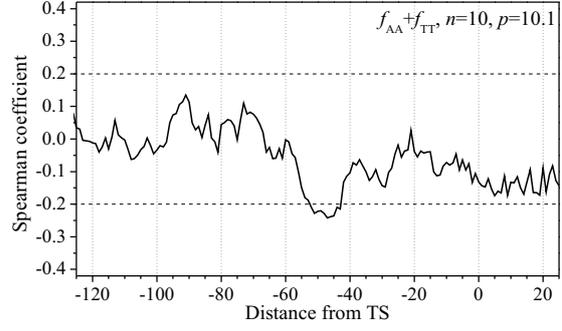
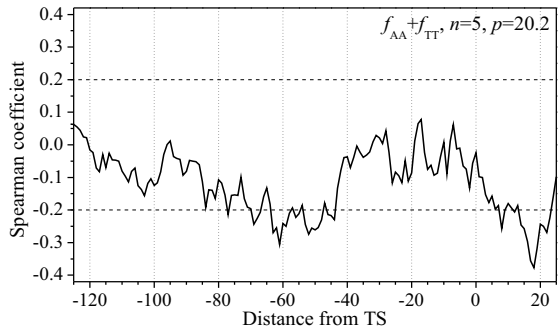
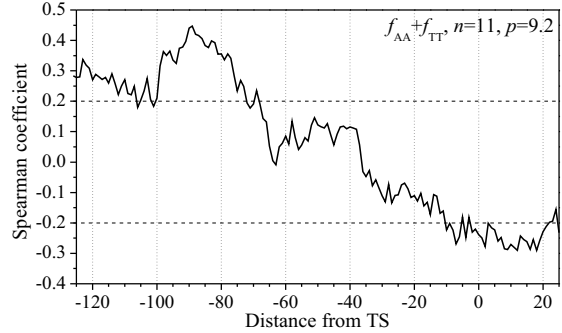

**B** Relaxation-repressed genes

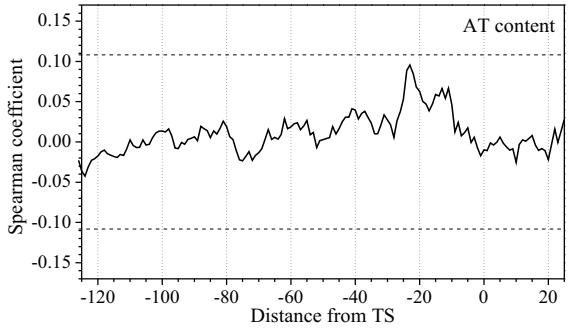
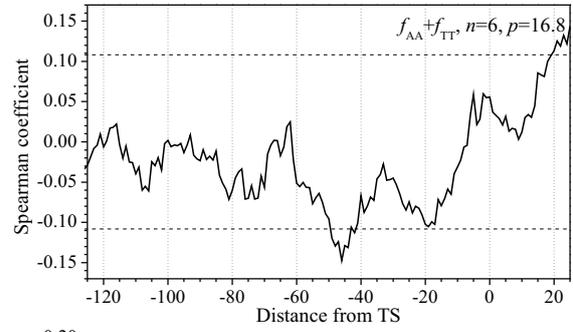
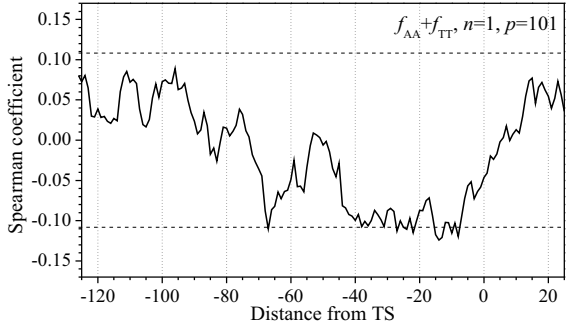
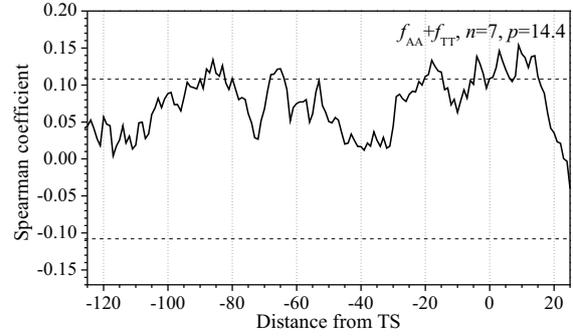
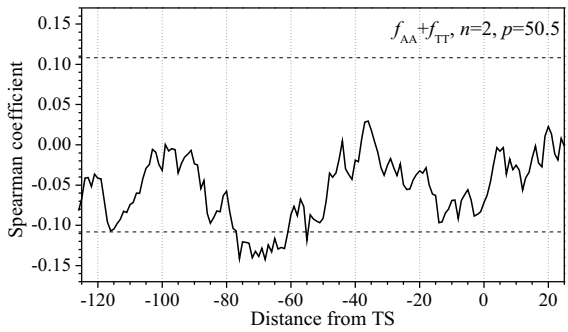
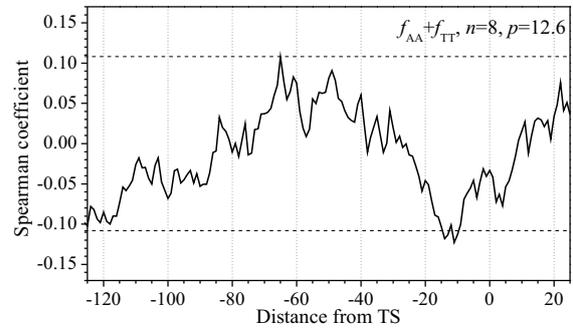
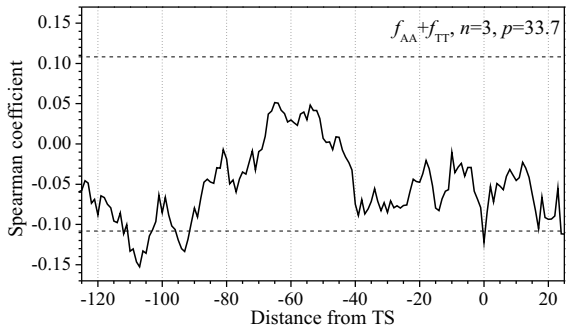
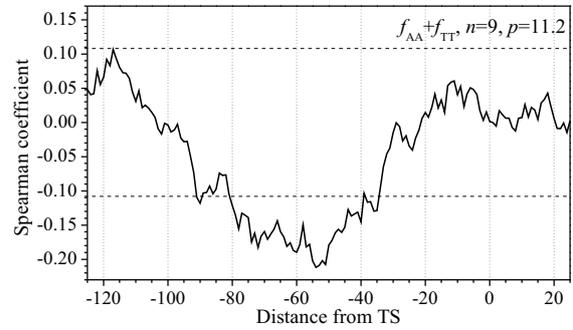
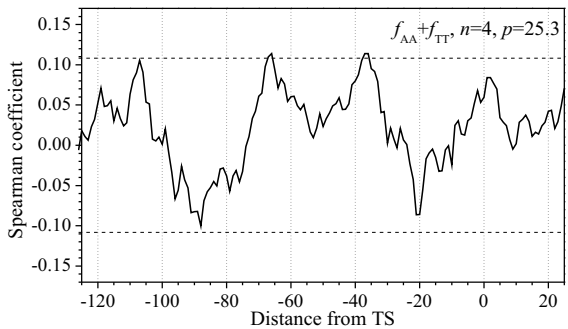
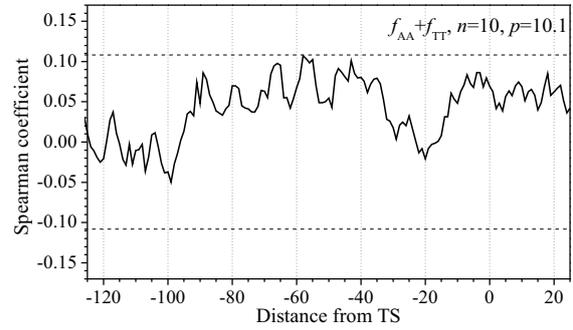
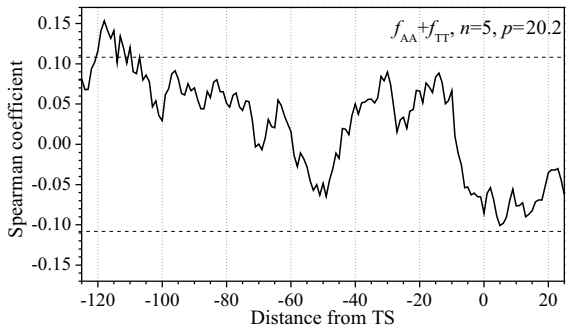
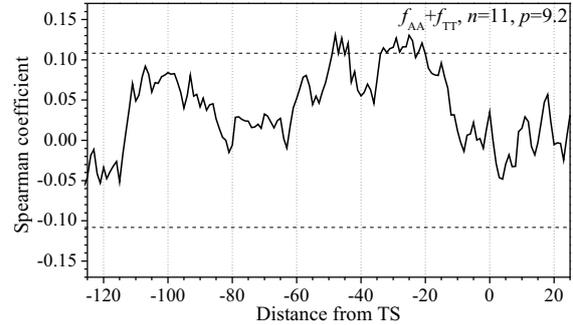

Reproducible dependence

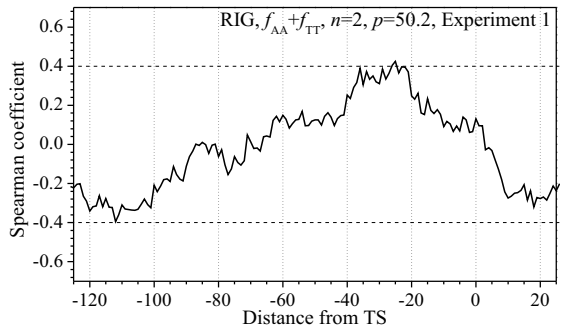
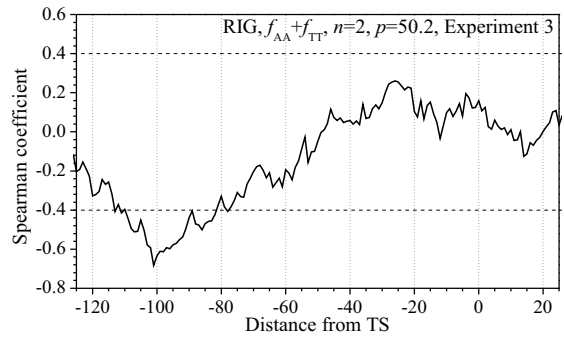
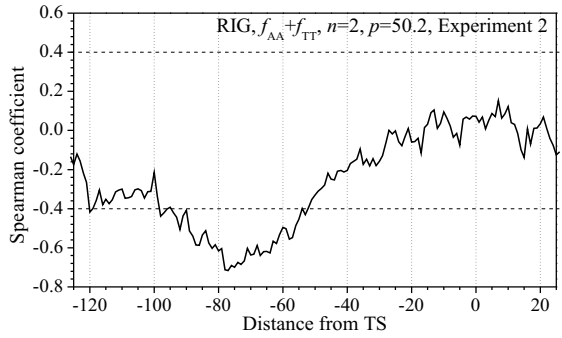
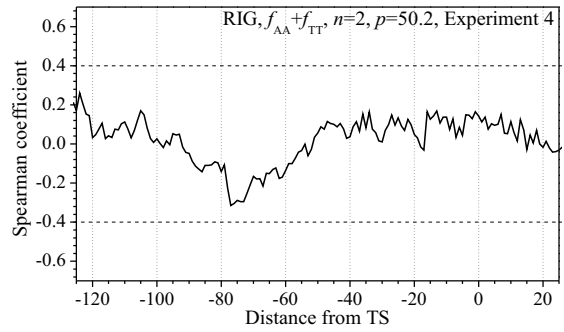
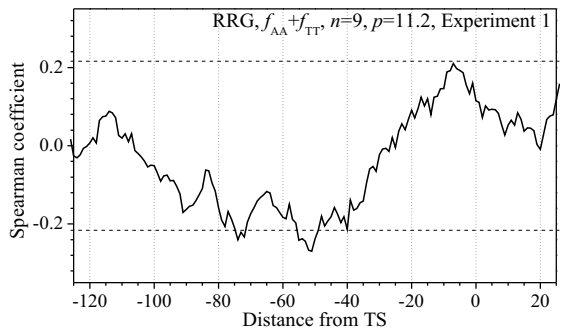
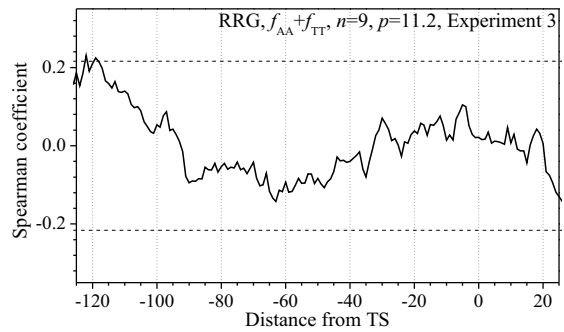
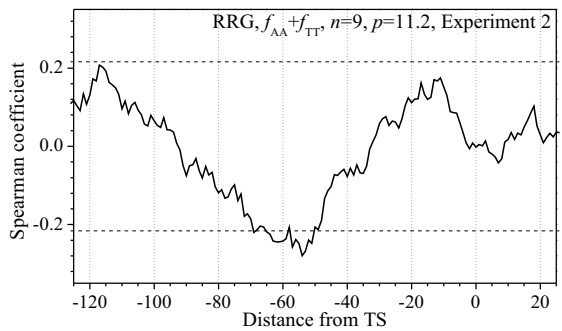
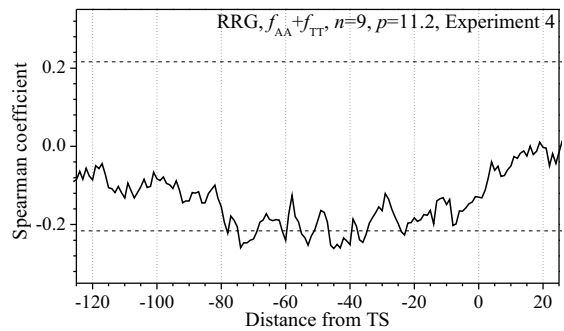
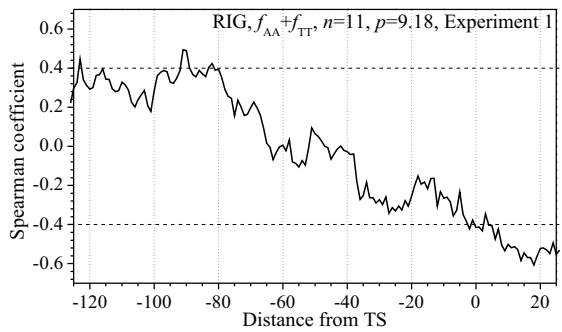
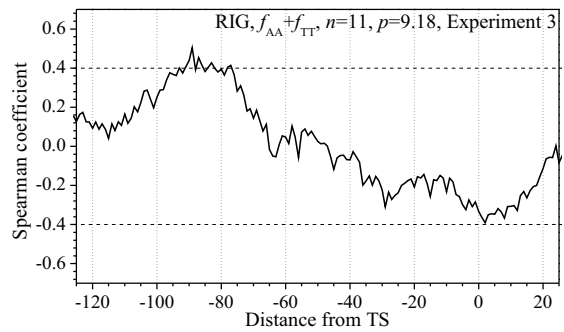
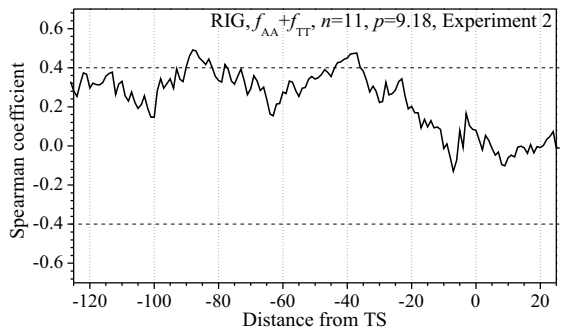
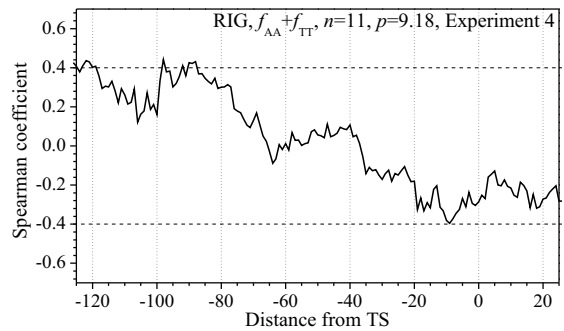

Non-reproducible dependence

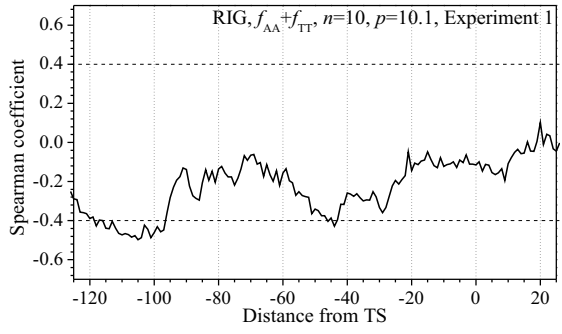
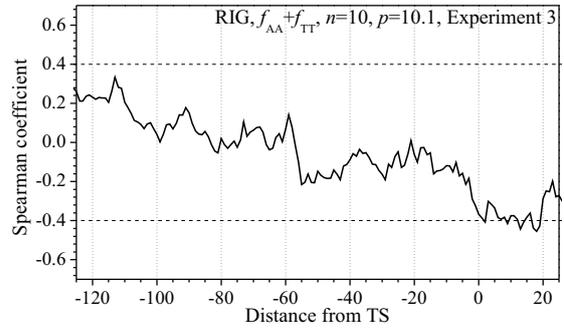
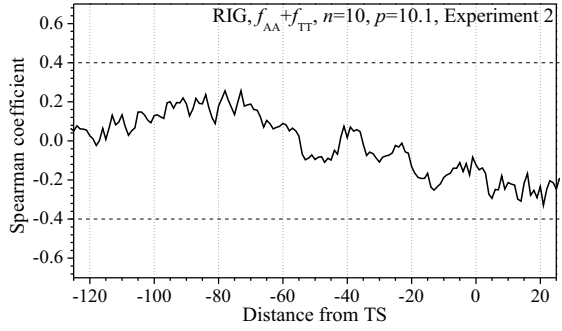
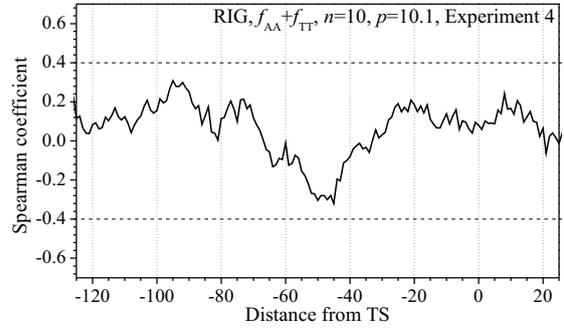

# Supplemental file 3

**Dependence of helical periodicity intensity on period in promoter sequences**

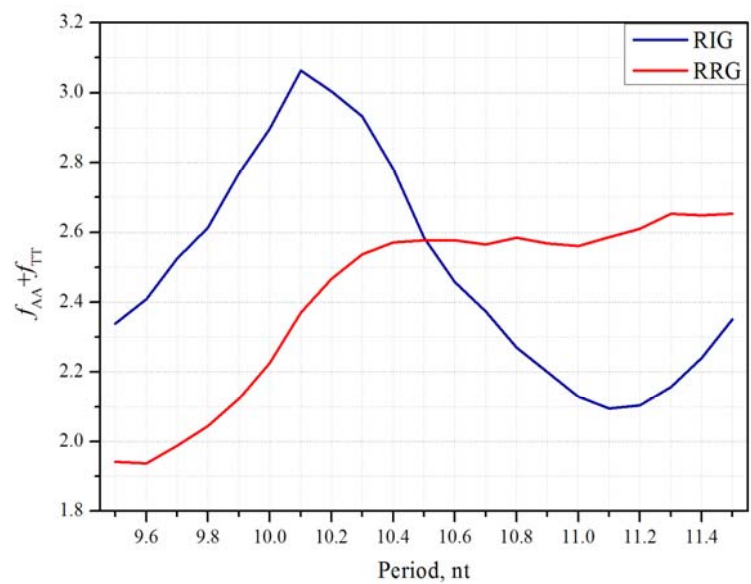

Dependence of periodicity intensity on period (resonance plot) was assessed by the sum of structure factors $f_{AA}+f_{TT}$ for harmonic $n=10$ averaged over corresponding promoter set and by varying the windows in the range $w=95–115$. 5'-end of all windows was fixed at position $−75$ from transcription start, whereas position of 3'-end varied.

**Dependence of helical periodicity intensity on period in the vicinity of promoters**

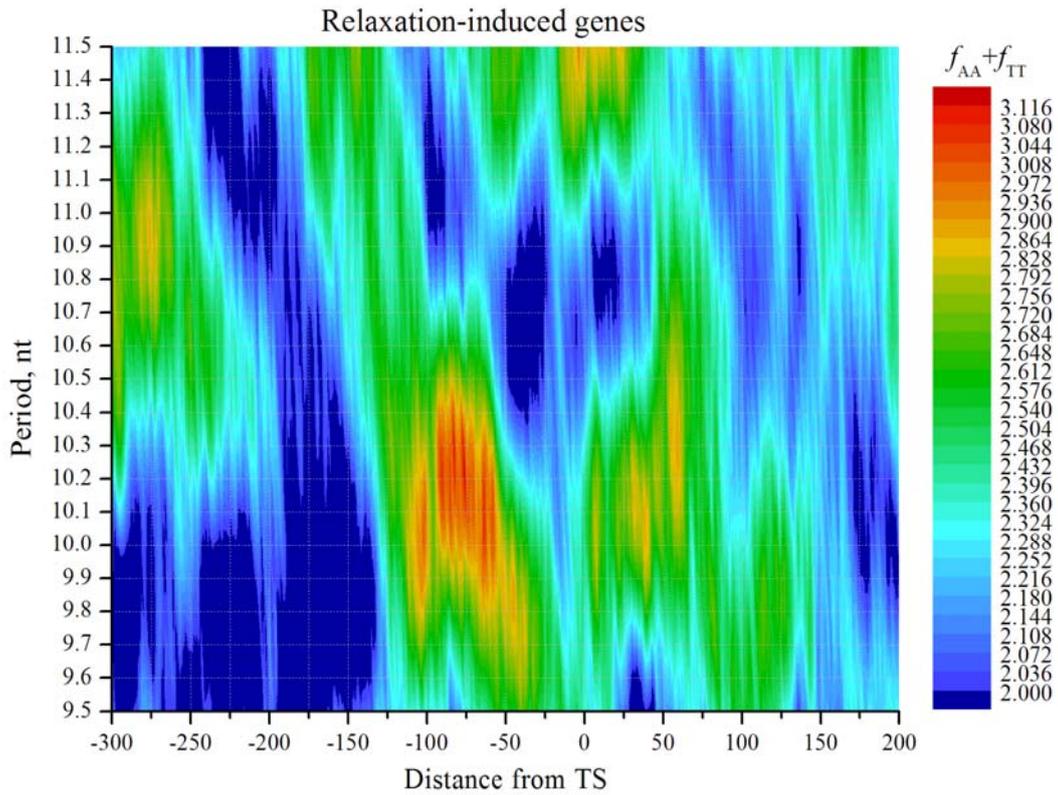

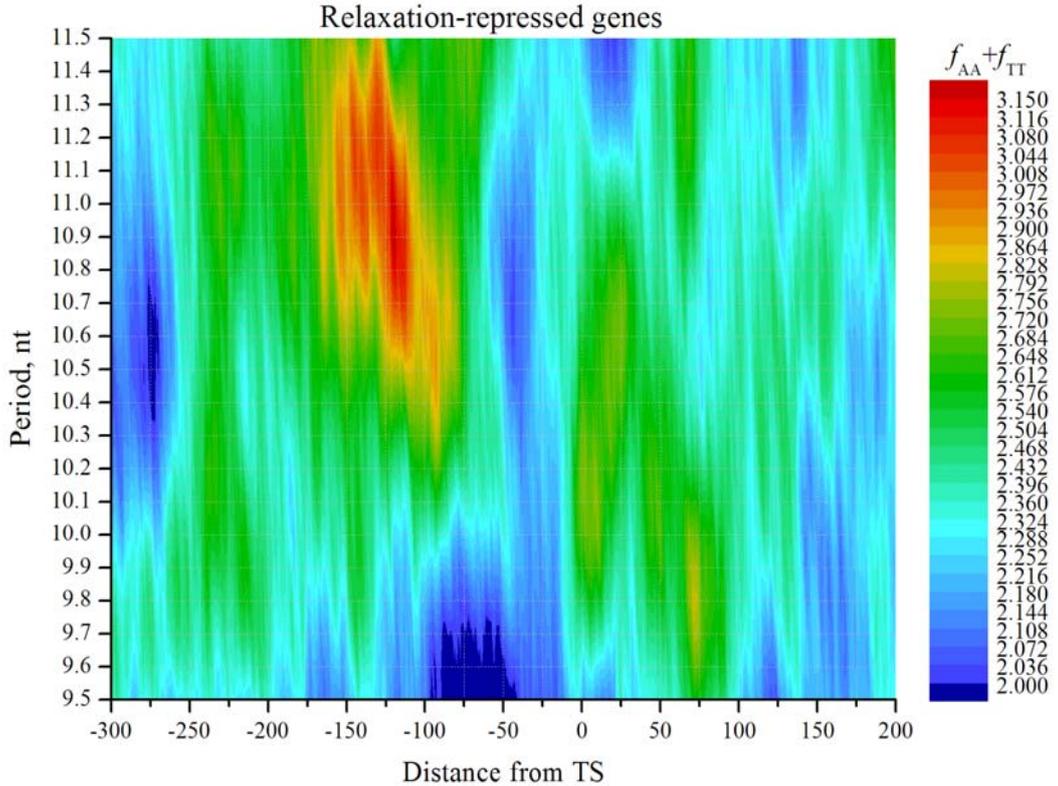

Dependence of periodicity intensity on period (resonance plot) was assessed by the sum of structure factors $f_{AA}+f_{TT}$ for harmonic $n=10$ and by varying the windows in the range $w=95–115$. The position is measured by the distance from the 5'-end of a sliding window to the transcription start (TS). For windows of fixed length, averaging is performed over all sliding windows at given position from TS.

**Normalized difference between spectra for RIG and RRG sets**

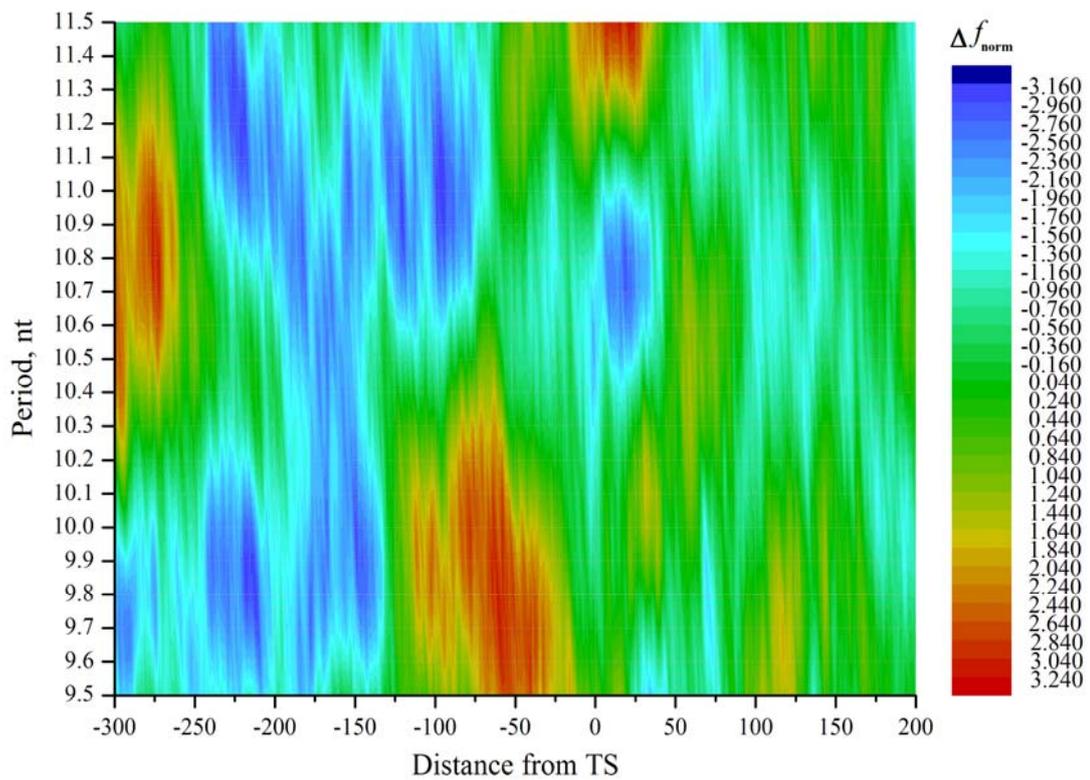

The normalized difference between spectra for helical periodicities in the vicinity of RIG and RRG promoters is defined by Eq. (5) (Materials and Methods). Note the statistically significant (by the criterion three standard deviations) change of sign from positive to negative in this difference at the transition from periodicities ~10 nt to ~11 nt in the vicinity of promoters (range from -100 to -25). This means that periodicities about 10 nt are characteristic of RIG promoters, whereas periodicities about 11 nt are characteristic of RRG promoters. Such relationship persists in the vicinity of promoters.

**Spearman correlations between expression log-ratios and periodicity intensity in promoter sequences**

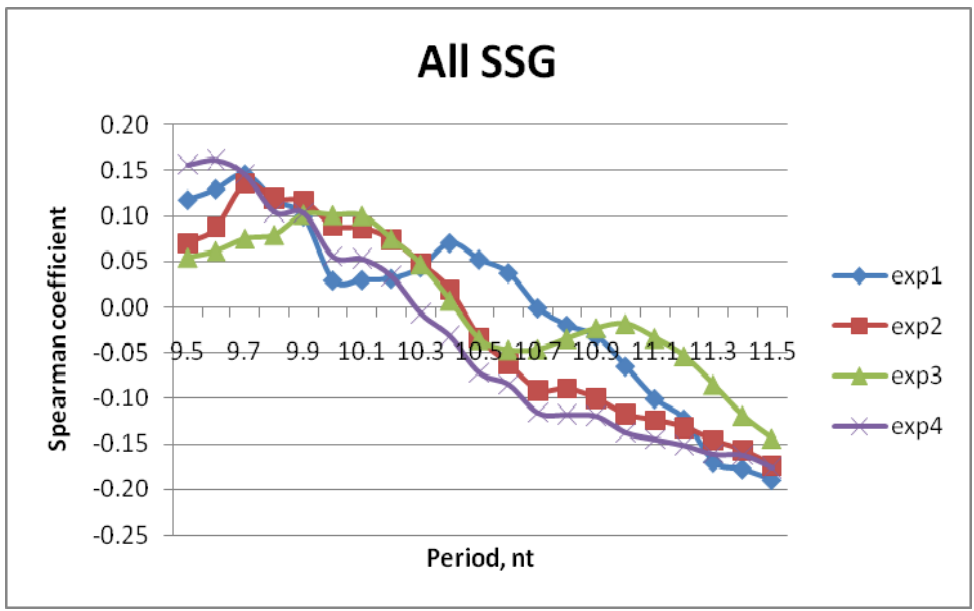

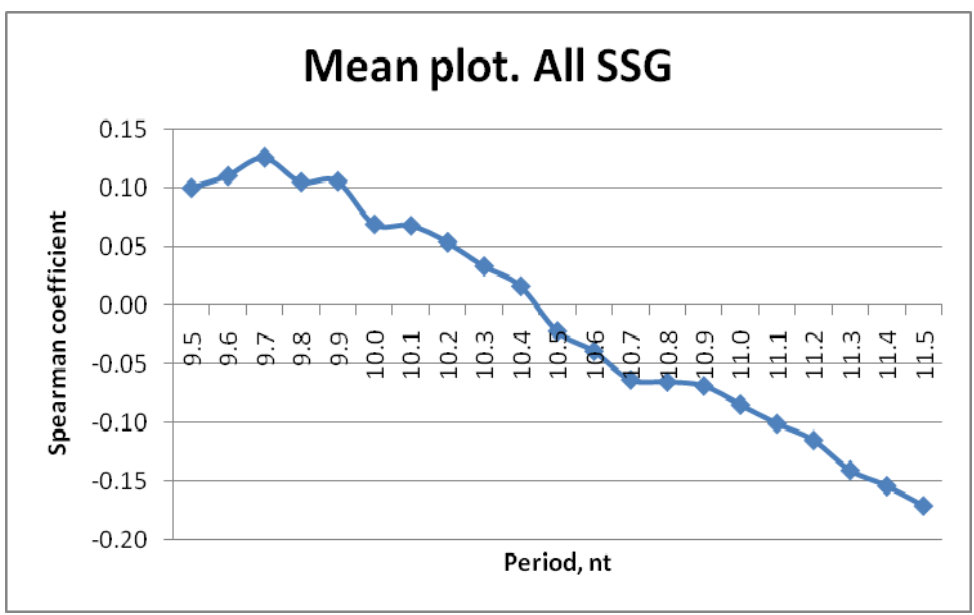

Periodicity intensity was assessed via sum of structure factors $f_{AA}+f_{TT}$ for harmonic $n=10$ and window range $w=95$-$115$. 5'-end of all windows was fixed at position $-75$ from transcription start, whereas position of 3'-end varied. The RIG and RRG promoters were united into one set.